\documentclass[a4paper,11pt]{article}
\usepackage{pos}
\usepackage{comment}

\def\rms{{\rm s}}

\def\sl{\rms_\lambda}

\def\proof{\noindent{\sl Proof:}\kern0.6em}

\def\dual{\mathstrut^*\kern-0.1em}

\def\lvec#1{\setbox0=\hbox{$#1$}
    \setbox1=\hbox{$\scriptstyle\leftarrow$}
    #1\kern-\wd0\smash{
    \raise\ht0\hbox{$\raise1pt\hbox{$\scriptstyle\leftarrow$}$}}
    \kern-\wd1\kern\wd0}
\def\rvec#1{\setbox0=\hbox{$#1$}
    \setbox1=\hbox{$\scriptstyle\rightarrow$}
    #1\kern-\wd0\smash{
    \raise\ht0\hbox{$\raise1pt\hbox{$\scriptstyle\rightarrow$}$}}
    \kern-\wd1\kern\wd0}
\def\slash#1{\setbox0=\hbox{$#1$}\setbox1=\hbox{$\kern1pt/$}
    #1\kern-\wd0\kern1pt/\kern-\wd1\kern\wd0}

\def\nabstar#1{{\nabla\kern0.5pt\smash{\raise 4.5pt\hbox{$\ast$}}
               \kern-5.5pt_{#1}}}

\def\nabbarstar#1{{\overleftarrow{\nabla}\kern0.5pt\smash{\raise 4.5pt\hbox{$\ast$}}
               \kern-5.5pt_{#1}}}

\def\nabdbarstar#1{{\overleftrightarrow{\nabla}\kern0.5pt\smash{\raise 4.5pt\hbox{$\ast$}}
               \kern-5.5pt_{#1}}}

\def\drvstar#1{{\partial\kern0.5pt\smash{\raise 4.5pt\hbox{$\ast$}}
               \kern-6.0pt_{#1}}}

\def\ldrvstar#1{{\lvec{\,\partial}\kern-0.5pt\smash{\raise 4.5pt\hbox{$\ast$}}
               \kern-5.0pt_{#1}}}

\def\MSbar{\overline{\rm MS\kern-0.5pt}\kern0.5pt}

\def\psibar{\overline{\psi}}

\def\zetabar{\bar{\zeta}}
\def\zetaprime{\zeta\kern1pt'}
\def\zetabarprime{\zetabar\kern1pt'}

\def\diracstar#1#2{
    \setbox0=\hbox{$\gamma$}\setbox1=\hbox{$\gamma_{#1}$}
    \gamma_{#1}\kern-\wd1\kern\wd0
    \smash{\raise4.5pt\hbox{$\scriptstyle#2$}}}

\def\Ds{D_{\rm s}}
\def\DsdagDs{\Ds{\Ds}^{\kern-1pt\dagger}}

\def\avg#1{{\kern1.0pt\overline{\kern-1.0pt#1\kern-1.0pt}\kern1.0pt}}

\newcommand{\be}{\begin{equation}}
\newcommand{\ee}{\end{equation}}
\newcommand{\bea}{\begin{eqnarray}}
\newcommand{\eea}{\end{eqnarray}}

\newcommand{\id}{1\!\!1}

\newcommand{\msbar}{{\rm \overline{MS\kern-0.05em}\kern0.05em}}

\newcommand{\ba}{\begin{eqnarray}}
\newcommand{\ea}{\end{eqnarray}}

\title{Computation of QCD meson screening masses at high temperature}

\author*[b,c]{Davide Laudicina}
\author[a]{Mattia Dalla Brida}
\author[b,c]{Leonardo Giusti}
\author[d]{Tim Harris}
\author[c]{Michele Pepe}

\affiliation[a]{Theoretical Physics Department, CERN,\\
CH-1211 Geneva 23, Switzerland}

\affiliation[b]{Dipartimento di Fisica, Università di Milano-Bicocca,\\
Piazza della Scienza 3, I-20126, Milano, Italy}

\affiliation[c]{INFN, sezione di Milano-Bicocca,\\
Piazza della Scienza 3, I-20126, Milano, Italy}

\affiliation[d]{School of Physics and Astronomy, University of Edinburgh,\\
Edinburgh EH9 3JZ, UK}

\emailAdd{davide.laudicina@mib.infn.it}
\emailAdd{mattia.dalla.brida@cern.ch}
\emailAdd{leonardo.giusti@mib.infn.it}
\emailAdd{tharris@ed.ac.uk}
\emailAdd{michele.pepe@mib.infn.it}

\abstract{We compute flavor non-singlet meson screening masses in the chiral limit of QCD with $N_f=3$ quarks. The calculation is
carried out at 12 temperatures from $T\approx 1$ GeV up to the electroweak scale. For each temperature we simulated
several lattice spacings, so as to be able to perform the continuum limit extrapolation with confidence at a few permille accuracy.
In the entire range of temperatures explored, the meson screening masses deviate from the  free theory result $2\pi T$ by at most a 
few percent. Their values, however, cannot be explained by one-loop perturbation theory up to the electroweak scale, where the
pseudoscalar and the vector screening masses are still significantly different within our precision. Chiral symmetry
restoration manifests itself through the degeneracy of the pseudoscalar and the scalar channels and of the vector and the axial ones.\\\\
\begin{flushright}
CERN-TH-2021-206
\end{flushright}
}

\FullConference{%
 The 38th International Symposium on Lattice Field Theory, LATTICE2021
  26th-30th July, 2021
  Zoom/Gather@Massachusetts Institute of Technology
}

\begin{document}
\maketitle

\section{Introduction}
QCD at finite temperature plays a crucial r$\hat{\rm{o}}$le in many fields of research, from the interpretation of experimental results from relativistic heavy ion colliders to cosmology and astrophysics. At high temperature, the low-energy scale $\Lambda_{\rm{QCD}}$ becomes less relevant as the temperature increases and the only relevant scale at asymptotically large temperatures is $\sim T$. At high temperatures the running coupling is small and the scale hierarchy $g^2 T\ll gT \ll T$ arises. The lattice represents the only theoretical framework in which the theory can be studied non-perturbatively, however up to now most of the works on the lattice are restricted to temperatures of about $T\approx 1$ GeV or below. The aim of this study is to design a strategy to study QCD non-perturbatively, up to very high temperatures, i.e. at the level of the electroweak scale or so. An analogous strategy has been successfully developed in the case of the pure $SU(3)$ gauge theory and allowed a precise determination of the Equation of State up to two orders of magnitude the critical temperature \cite{Giusti:2016iqr}. As a first concrete application of this strategy we present our results on the calculation of the QCD non-singlet meson screening masses \cite{Work}.
\section{The QCD meson screening masses}
In the framework of thermal QCD maybe the simplest class of observables one can study are those obtained from two-point spatial correlation functions of fermionic bilinears. The large-distance behaviour of these correlation functions is dominated by the so called screening masses, namely the inverse of the correlation lengths, which describe the response of the quark and gluon plasma when a meson is put in the system. In this work we are interested in flavor non-singlet bilinear operators
\begin{equation}
{\cal O}^a (x) = \psibar(x) \Gamma_{{\cal O}} \, T^a \,\psi(x)\;,\\
\end{equation}
where $T^a$ are the traceless generators of $SU(3)$ which describe the structure of the operator in flavor space, $\Gamma_{{\cal O}}=\left\{\id,\gamma_5,\gamma_\mu,\gamma_\mu\gamma_5\right\}$ and the corresponding operators are named ${\cal O} = \left\{S,P,V_{\mu},A_{\mu}\right\}$. The spatially separated two-point correlation functions of these operators can be defined in the continuum as 
\begin{equation}\label{eq:2pt}
  C_{{\cal O}}(x_3) =\int dx_0 dx_1\, dx_2\, \langle {\cal O}^a (x) {\cal O}^a (0) \rangle\;,
\end{equation}
where no summation over $a$ is understood. The screening masses are then defined as 
\begin{equation}
m_{{\cal O}} = - \lim_{x_3\rightarrow\infty} \frac{d}{d x_3} \ln\Big[C_{{\cal O}}(x_3)\Big]\; . 
\end{equation}
To date the QCD meson screening masses have been computed in perturbation theory and on the lattice. On the one hand, the lattice calculations are restricted to temperatures below $T\approx 1$ GeV \cite{Bazavov}, on the other hand perturbation theory is expected to be reliable at asymptotically large temperatures. In this regime the theory becomes static, the degrees of freedom in the temporal extent can be integrated out and the calculation can be carried out in the resulting dimensional-reduced effective theory. In the 3d theory the quark fields develop a thermal mass of about $\pi T$ and are treated as heavy, static fields \cite{Laine}. In this approximation, independently of the flavor structure of the interpolating operators, the meson non-singlet screening masses at one-loop are
\begin{equation}
m^{\rm PT}_{{\cal O}} = 2 \pi T + \frac{g_E^2(T)}{3\pi} \big( 1 + 0.93878278 \big)\,=\, 2\pi T\left(1+0.032739961\cdot g^2\right)\; ,
\label{eq:pt}
\end{equation}
where $g_{E}^2$ is the effective coupling of the dimensional-reduced theory as defined in Ref. \cite{Laine} and its value is fixed by matching the 3d theory with QCD.
\section{Lattice setup}
We regularized the theory on a 4-dimensional lattice with compact temporal extent $L_0$. All the fields are taken to be periodic in space while in the temporal direction they satisfy shifted boundary conditions \cite{Giusti1, Giusti2, Giusti3, DallaBrida:2020gux}.

We simulated 12 values of temperature $T_0,...,T_{11}$ ranging from 1 GeV up to 160 GeV. All the simulations were performed using the Hybrid Monte Carlo algorithm in the presence of $N_f=3$ flavors of quarks in the chiral limit. We discretized the fermionic sector of the action considering the $O(a)$-improved Wilson-Dirac operator. The gauge sector was discretized, at the highest temperatures $T_0,...,T_8$, with the Wilson plaquette action while at $T_9,T_{10}$ and $T_{11}$ we considered the tree-level Symanzik improved action. The strategy proposed in this work profits from the non-perturbative determination of the QCD running coupling which was used to fix the lines of constant physics. In particular, these have been fixed through the Schr\"odinger functional (SF) finite volume coupling for $T_0,...,T_8$ and the gradient flow (GF) coupling for $T_9,T_{10}$ and $T_{11}$ \cite{alpha3,alpha6}. Once the lines of constant physics have been fixed, the critical mass is obtained by requiring the PCAC mass to vanish in a finite volume with SF boundary conditions.

The simulations were performed considering several lattice spacings so as to be able to perform the continuum limit extrapolation with confidence. In particular, at $T_1,...,T_8$ we simulated 4 different lattice spacings ($L_0/a=4,6,8,10$), at $T_9,...,T_{11}$ we simulated 3 lattice spacings ($L_0/a=4,6,8$) and at the highest temperature $T_0\approx 160$ GeV, 2 different lattice spacings have been simulated ($L_0/a=4,6$).

Finite volume effects are kept under control by simulating large lattices. We considered lattices with $L/a=288$ in each spatial extent so as to keep $LT$ between 20 and 50 for each simulation.
\section{Numerical study}
In the calculation of the screening correlator in eq.~\eqref{eq:2pt} the inversion of the Dirac operator is required in order to estimate the quark propagator. As mentioned, at high temperature the quark fields develop a thermal mass of about $\pi T$ which provides an infrared cutoff to the quark propagator. This leads, as a consequence, to the fact that for large source-sink separations $|x-y|$ the quark propagator $D^{-1}(x,y)$ is extremely suppressed. We have solved this problem by using a distance-preconditioned version of the Dirac equation in which we factorized out the bulk of the exponentially-suppressed behaviour of the quark propagator \cite{deDivitiis}. We then solved the distance-preconditioned equation
\begin{equation}
\tilde{D}\tilde{\psi}=\tilde{\eta}
\end{equation}
where
\begin{equation}
\tilde{D}=M^{-1}DM\,, \quad \tilde{\psi}=M^{-1}\psi\,, \quad \tilde{\eta}=M^{-1}\eta
\end{equation}
with the preconditioning matrix $M$ defined  as
\begin{equation}
M(x_3,y_3)=\cosh\left\{ m_{M}(x_3-y_3-L/2) \right\} 
\end{equation}
where $m_{M}$ has been tuned for the different lattice spacings so as to have $m_{M}\approx \pi/(\sqrt{2} L_0)$.

\begin{figure}[t!]
  \begin{center}
\includegraphics[width=0.9\textwidth]{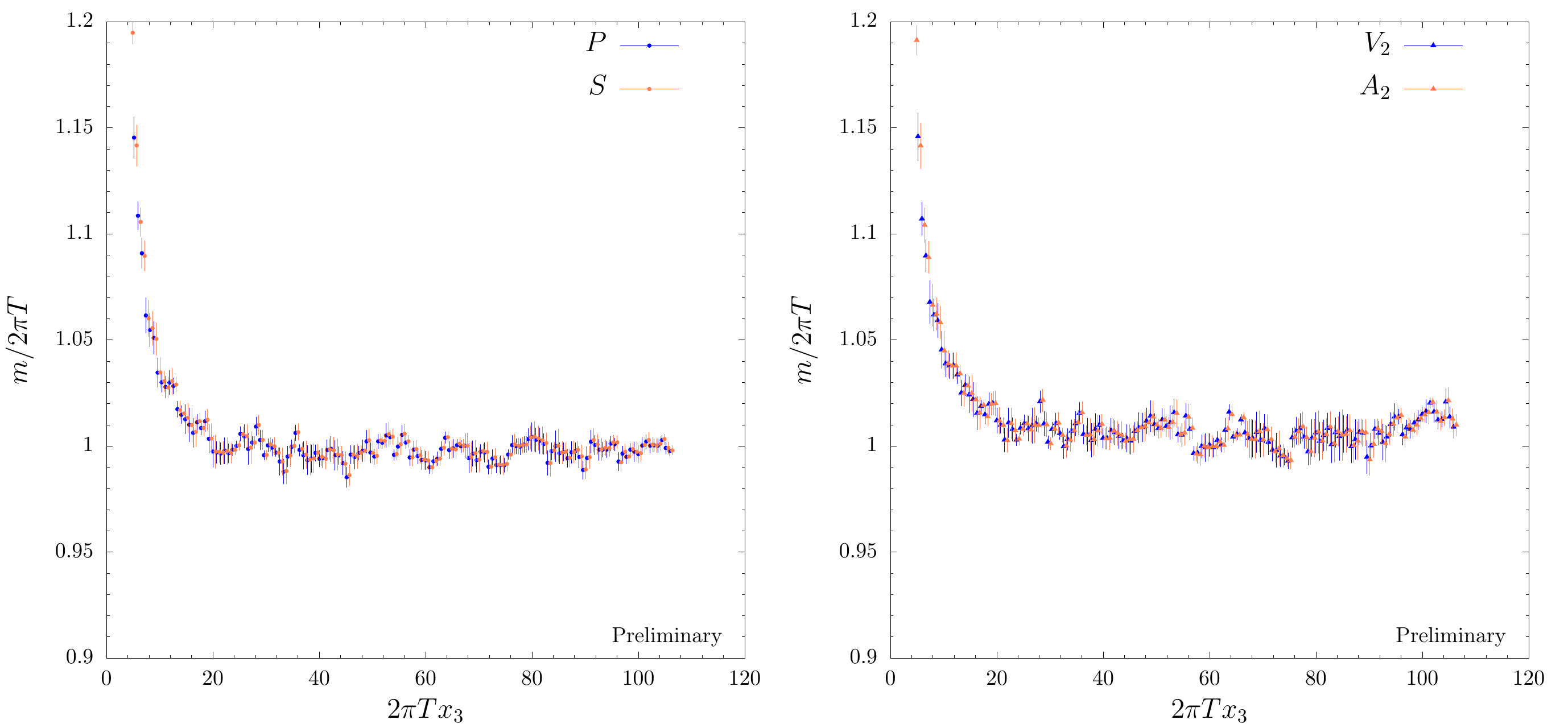}
\caption{The effective mass versus the separation on the lattice for the pseudoscalar and the scalar channel (left panel) and for vector and the axial (right panel) at the physical temperature $T_3=33$ GeV. The scalar and the axial datasets are shifted by a factor of $0.5$ to the right for readability\label{fig:chiral}.}
\end{center}
\end{figure}

Once the screening correlator has been computed, the screening mass on the lattice is extracted by taking the large separation limit of the screening correlation function and its expression can be written as
\begin{equation}
m_{{\cal O}} (x_3)= \frac{1}{a} {\rm arcosh}\Big[\frac{C_{{\cal O}}(x_3+a)+C_{{\cal O}}(x_3-a)}{2\, C_{{\cal O}}(x_3)}\Big]\; . 
\end{equation}
In the whole range of temperatures explored we observe the complete degeneracy between the pseudoscalar and the scalar masses and between the axial and the vector ones. In fig. \ref{fig:chiral}, we show an example of the effective mass at the physical temperature $T_3=33$ GeV, after having symmetrized the screening correlator with respect to $x_3=L/2$, for the pseudoscalar and the scalar channel (left panel) and for the vector and the axial one (right panel). For all the other temperatures we obtained similar plots. The degeneracy of these masses is a clear indication that chiral symmetry is effectively restored in the entire range of temperature. For this reason in the following discussion we restrict ourselves to the pseudoscalar and to the vector screening masses.
\begin{figure}[t!]
  \begin{center}
\includegraphics[width=0.9\textwidth]{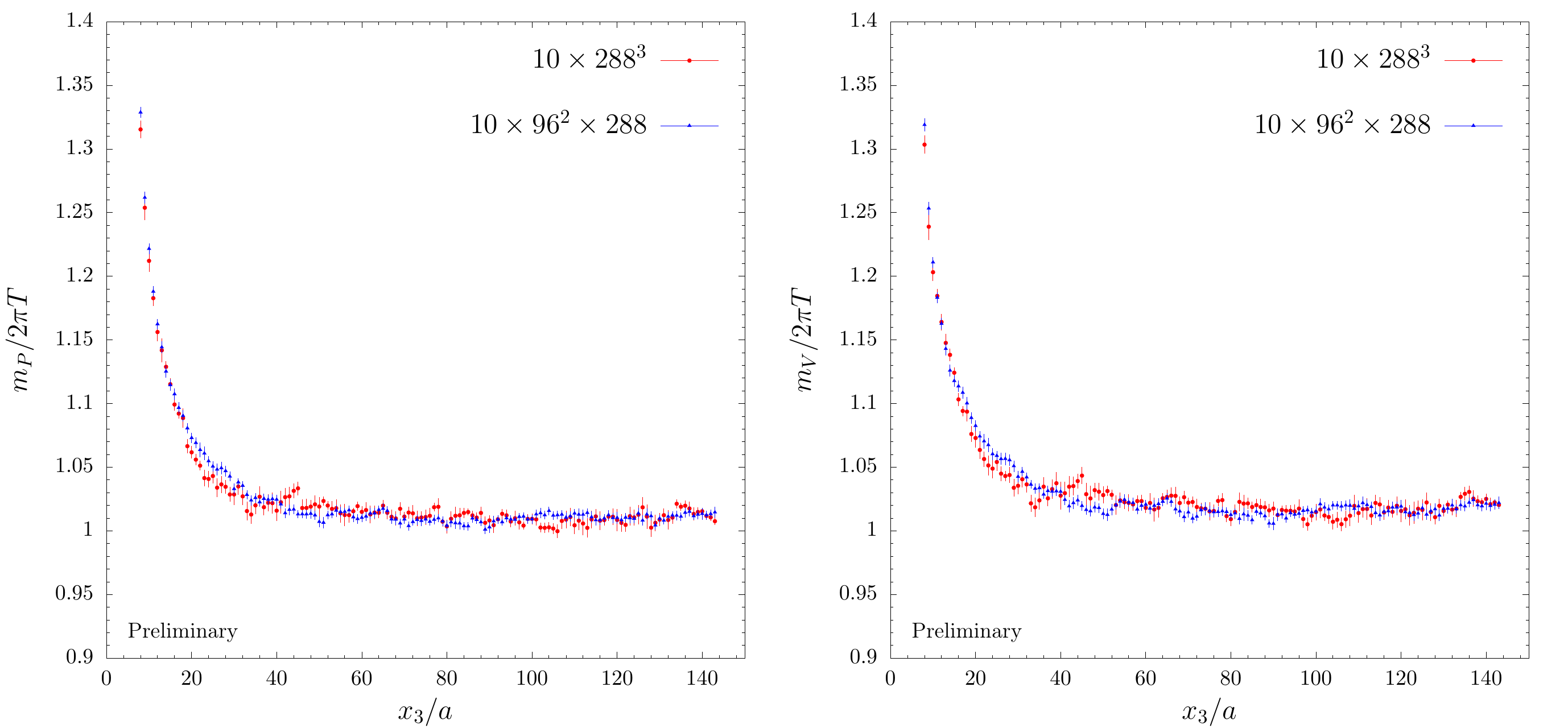}
\caption{Plot of the effective mass function for the pseudoscalar and vector correlators at the physical temperature $T_1=82$ GeV. The red points are the results in a volume with 288 sites in each spatial direction, while the blue ones in a spatial volume of $96^2\times 288$ sites\label{fig:plateau}.}
\end{center}
\end{figure}

The analysis of finite volume effects has been carried out at $T_0$, $T_1$ and $T_{11}$ on the volumes corresponding to $L_0/a=6$, $L_0/a=10$ and $L_0/a=8$ respectively, in order to check that finite size effects are negligible in the entire range of temperatures explored. As we can see from, e.g. fig. \ref{fig:plateau}, both for the pseudoscalar and for the vector channel finite volume effects are under control and negligible within our numerical precision.

\begin{figure}[t!]
\begin{center}
\includegraphics[width=0.45\textwidth]{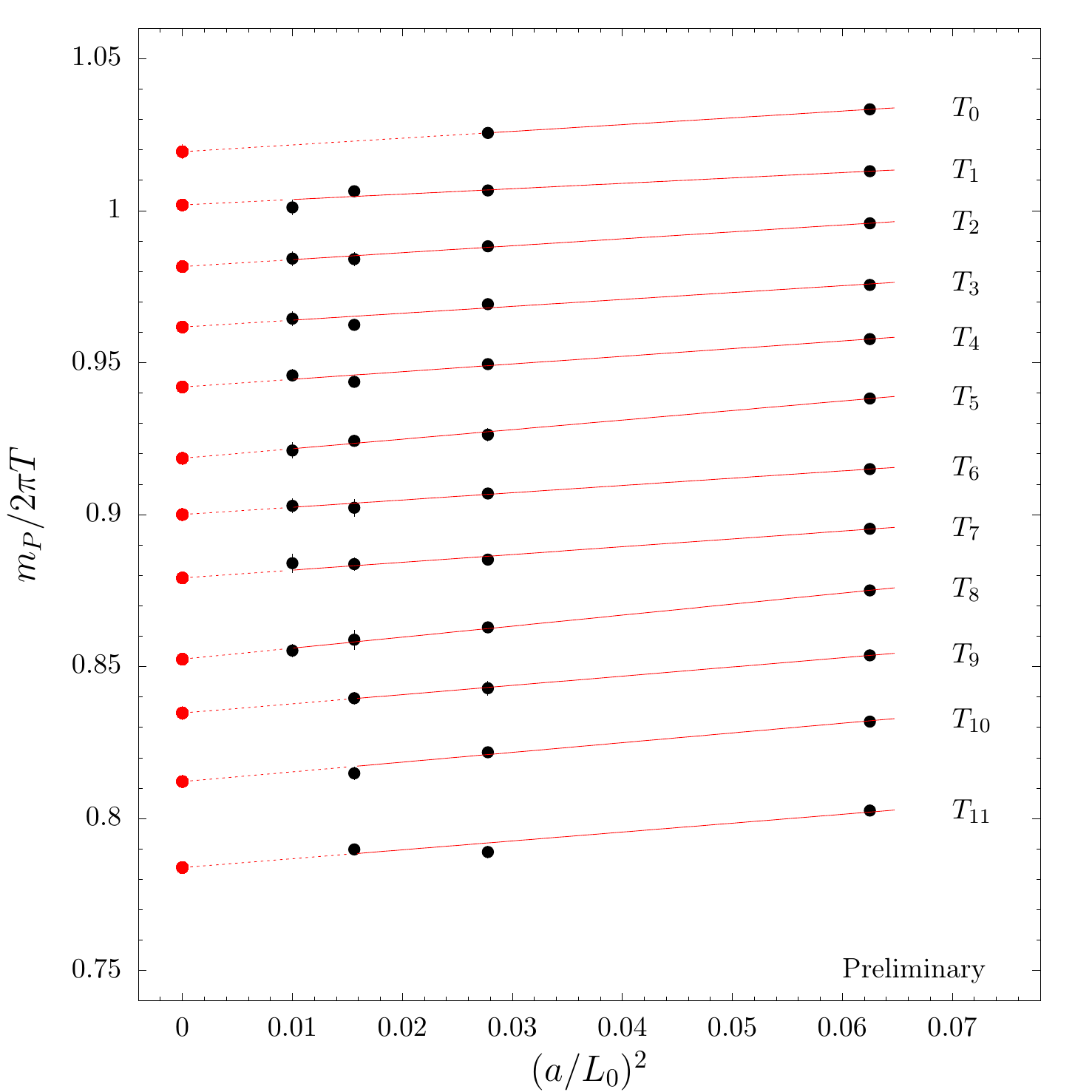}
\includegraphics[width=0.45\textwidth]{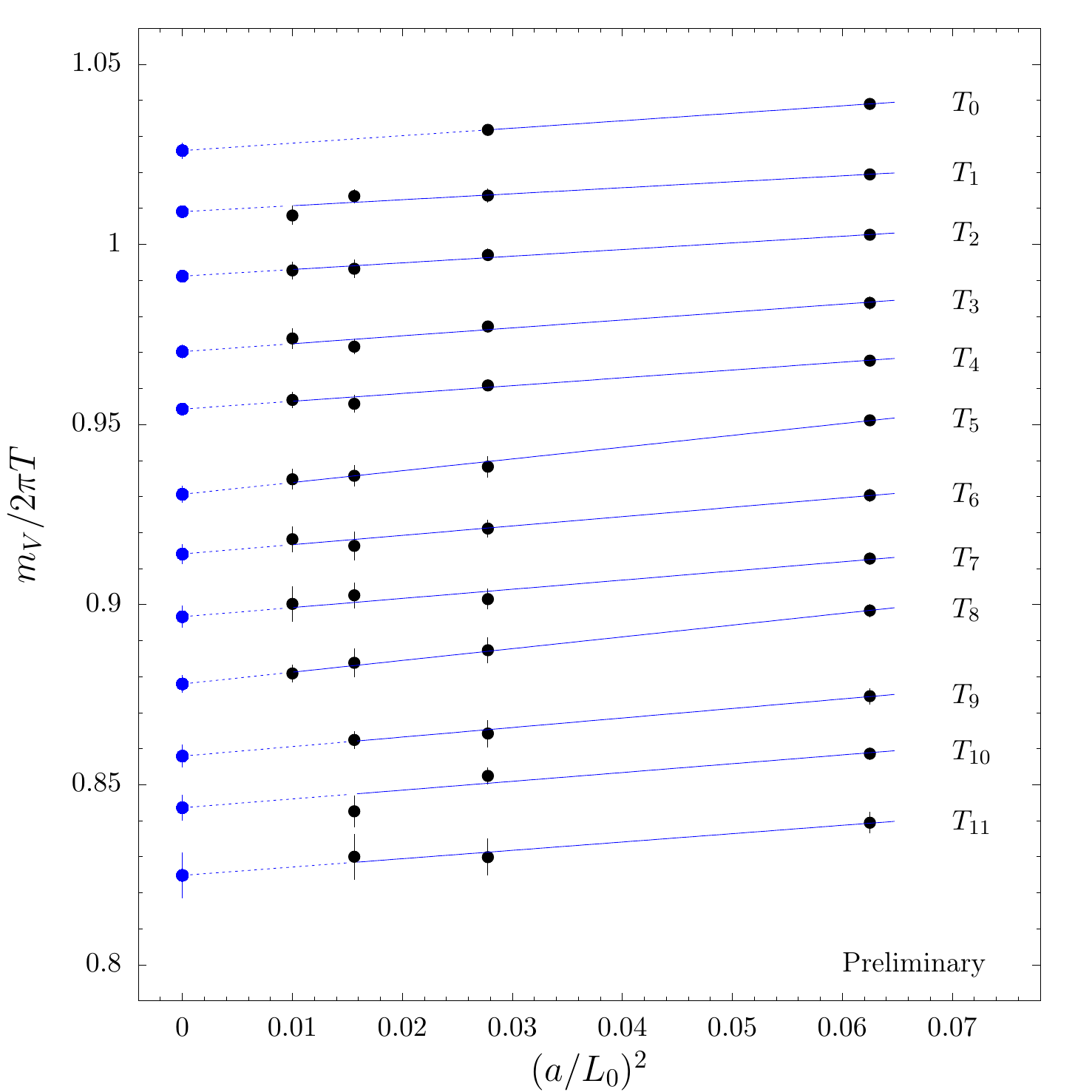}
  \caption{Numerical results of the tree-level improved pseudoscalar (left) and vector (right) screening mass at finite lattice spacing (black dots). The red and the blue lines represent the linear extrapolations in $(a/L_0)^2$ to the continuum limit (red and blue points); each temperature is analyzed independently on the others. The extrapolation for a given temperature $T_i$ is shifted by $0.02\cdot i$ downward for readability.
\label{fig:PV_extrCL}}
\end{center}
\end{figure}

In order to remove the dependence on the lattice spacing we performed the continuum limit extrapolation. To speed up the convergence to the continuum we computed analytically the free theory contribution to the screening masses on the lattice. Then we give the tree-level improved definition of the screening masses at finite lattice spacing
\begin{equation}
\label{eq:tli}
m_{{\cal O}} \longrightarrow m_{{\cal O}} -
\Big[m^{\rm{free}}- 2\pi T\Big]\; , 
\end{equation}
where $m^{\rm{free}}$, which is the mass in the free lattice theory computed in the infinite volume limit, is independent of the flavor structure of the operator. This procedure allowed us to perform the continuum limit extrapolation with confidence at a few permille accuracy. In fig. \ref{fig:PV_extrCL} we show the extrapolations for the pseudoscalar screening mass (left panel) and for the vector screening mass (right panel). We considered as fit ansatz a single correction proportional to $(a/L_0)^2$ for all the temperatures and each extrapolation is treated as independent on the others.
For $T_1,...,T_8$ we also performed the same kind of analysis by omitting in the extrapolation the results on the coarser lattice spacing ($L_0/a=4$). We fitted the data also including in the fit ansatz terms proportional to $(a/L_0)^2\ln(a/L_0)$ and to $(a/L_0)^4$. In all the cases the additional coefficients are compatible with zero and the intercepts are in excellent agreement with those of the previous fits. In the following we take as best estimates of the masses the results obtained by fitting all the data available for each temperature.

\begin{figure}[t!]
\begin{center}
\includegraphics[width=0.45\textwidth]{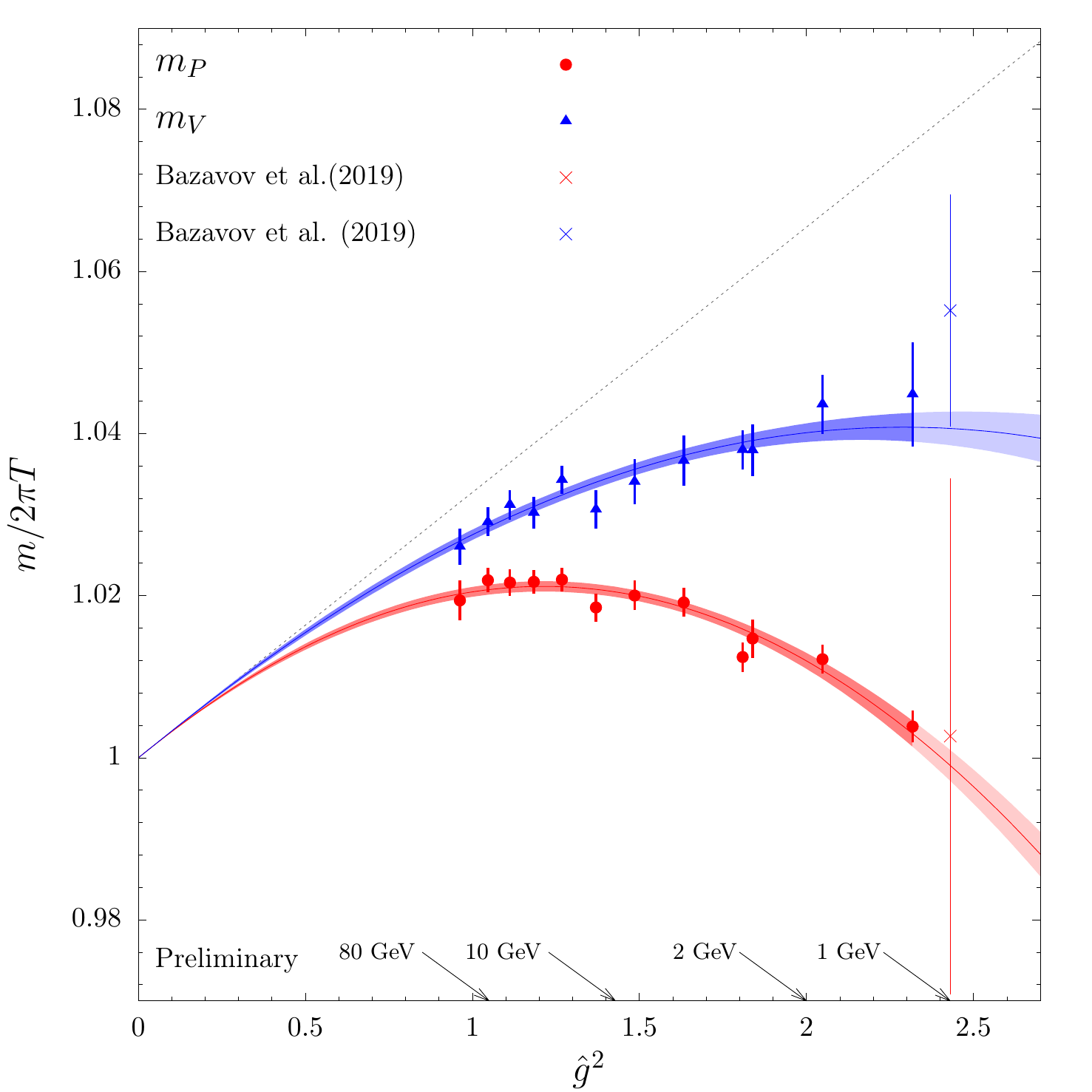}
\caption{Pseudoscalar (red) and vector (blue) screening masses versus $\hat{g}^2$. The curves represent the best fits. The dashed line is the analytically known contribution. The two points on the right hand side of the plot represent the result at $T=1$ GeV for the pseudoscalar and for the vector screening masses obtained in Ref. \cite{Bazavov}\label{fig:temp_dep}.}
\end{center}
\end{figure}

In fig. \ref{fig:temp_dep} we show the screening masses in the continuum limit, normalized to the free theory result $2\pi T$, as a function of the temperature. The temperature dependence is parametrized through the two-loop renormalized coupling in the $\MSbar$ scheme evaluated at the scale $\mu=2\pi T$, which expression reads
\begin{equation}
\hat{g}^{-2}(T) \, = \, \frac{9}{8\pi^2}\ln\left(\frac{2\pi T}{\Lambda_{\MSbar}}\right)+\frac{4}{9\pi^2}\ln\left[2\ln\left(\frac{2\pi T}{\Lambda_{\MSbar}}\right)\right]\, ,
\end{equation}
where $\Lambda_{\MSbar}=0.341$ GeV \cite{alpha1}.
We see how in the entire range of temperature the bulk of the screening masses is described by the free theory result plus a few percent positive deviation. In particular, we observe at most a 2\% positive deviation for the pseudoscalar mass and a 4\% deviation for the vector one. By analyzing the results in more detail we see how even at the highest temperatures we simulated the one-loop perturbative prediction is not satisfactory within our precision, since the pseudoscalar and the vector mass are still significantly different at the electroweak scale. By performing a fit in the temperature with a polynomial function in $\hat{g}^2$ we find that at much higher temperature than the electroweak scale the intercept and the linear coefficient of the extrapolation are compatible with the free theory result and with the first perturbative correction respectively. This is a clear indication that the one-loop perturbative result gives a reliable estimate of the screening masses only well above the electroweak scale. On the other hand, at low temperature one-loop perturbation theory becomes unreliable. In particular, this is evident by the fact that in the low temperature regime the pseudoscalar screening mass exhibits a negative slope in $\hat{g}^2$, while the perturbative calculation predicts a positive slope in the entire range of temperature.

\section{Conclusions}
In this work, as a first application of our strategy to study thermal QCD at extremely high temperature, we computed for the first time the QCD meson screening masses on the lattice in the range of temperature between $T\approx 1$ GeV and the electroweak scale. We simulated large volumes to keep finite volume effects under control and different lattice spacings so as to be able to perform the continuum limit extrapolation with confidence. This allowed us to determine the screening masses with a few permille accuracy in the continuum limit. In the entire range of temperature we observed chiral symmetry restoration due to the degeneracy between the pseudoscalar and the scalar screening masses as well as the vector and the axial ones. Our results show that the bulk of the masses is dominated by the free theory result with at most a 2\% positive deviation for the pseudoscalar screening mass and a 4\% for the vector one. However, even at the highest temperature we simulated our results cannot be explained by the one-loop perturbative result, since at the electroweak scale the pseudoscalar and the vector mass are still significantly different within our precision. In addition, in the low temperature regime we observe the failure of the one-loop perturbative prediction emphasized by the negative slope of the pseudoscalar mass in $\hat{g}^2$ for $T\lesssim 10$ GeV.

In conclusion, the strategy proposed in this work and the results achieved in the present study pave the way for a non-perturbative study of the fundamental properties of thermal QCD from low to high temperature.

\section*{Acknowledgement}
We acknowledge PRACE for awarding us access tothe HPC system MareNostrum4 at the Barcelona Supercomputing Center (Proposals n. 2018194651 and 2021240051) where most of the numerical results presented in this paper have been produced. We also thanks CINECA for providing us with computer-time on Marconi (CINECA- INFN, CINECA-Bicocca agreements, ISCRA B project HP10BF2OQT). The R\&D has been carried out on the PC clusters Wilson and Knuth at Milano-Bicocca. We thank all these institutions for the technical support. Finally we acknowledge partial support by the INFN project ``High performance data network''.

\end{document}